\documentclass[published]{JHEP3} 

\JHEP{00(2007)000}

\JHEPspecialurl{http://jhep.sissa.it/JOURNAL/JHEP3.tar.gz}

\usepackage{epsfig,multicol,bbm}

\newcommand\fverb{\setbox\fverbbox=\hbox\bgroup\verb}
\newcommand\fverbdo{\egroup\medskip\noindent%
            \fbox{\unhbox\fverbbox}\ }
\newcommand\fverbit{\egroup\item[\fbox{\unhbox\fverbbox}]}
\newbox\fverbbox

\newcommand{\newc}{\newcommand}
\newc{\beq}    {\begin{equation}}
\newc{\eeq}    {\end{equation}}

\newc{\beqa}    {\begin{eqnarray}}
\newc{\eeqa}    {\end{eqnarray}}
\newc{\bs}    {\section}
\newc{\no}    {\\ \nonumber}

\newc{\st}    {\stackrel}

\title{  Minimum mass of galaxies from BEC or scalar field dark matter}

\author{Jae-Weon Lee\\
 Department of energy resources development, Jungwon
 University,  5 dongburi, Goesan-eup, Goesan-gun Chungbuk, 367-805 Republic of  Korea
\\
E-mail:\email{scikid@gmail.com}}
    \author{Sooil Lim\\
 Department of Physics and Astronomy, Seoul National University, Seoul, 151-747, Republic of  Korea}

\received{\today}       
\accepted{\today}       

\abstract{
 Many problems of cold dark matter models  such as the cusp
   problem and the missing satellite problem can be alleviated, if
   galactic halo dark matter particles are  ultra-light scalar particles and
   in  Bose-Einstein condensate (BEC), thanks to a characteristic length scale
    of the  particles.
 We show that this finite length scale of the dark matter can also  explain  the recently observed
  common central mass
 of the Milky Way satellites ($\sim 10^7 M_\odot$) independent of their luminosity, if
 the mass of the dark matter particle is about $10^{-22} eV$.
}


\begin{document}


\section{Introduction}

Dark matter (DM)
is one of the most important
 puzzles in modern physics and cosmology.
 Since the presence of DM was  inferred
from gravitational effects on visible matter
in galaxies~\cite{DMreview},
 a good DM model should explain the structure and evolution of galaxies.
  The dwarf
 galaxies seem to be the smallest dark matter dominated astronomical objects and, hence, are
 ideal for studying the nature of  DM~\cite{gilmore}.
 Observational data suggest that a typical dwarf galaxy has a size never less than $O(10^2)~ pc$.
This could pose a challenge to the structure formation theory based on
 a simple cold dark matter (CDM) model~\cite{gilmore}, because
  there is no scale that can be easily accessed observationally in
the CDM theory. (However, there is surely  a scale in the structure formation of baryons in DM halos.
 Understanding theoretically the scale is a major topic
of present-day research.)

  Although
CDM such as WIMP( Weakly Interacting Massive Particles) with the cosmological constant ($\Lambda$CDM) model is popular and remarkably successful
 in explaining
 large scale structures of the
universe, it seems to encounter many other problems such as the cusp problem and the missing satellite
 problem on a scale of galactic or sub-galactic structures~\cite{DMmodels}. For example, numerical
studies with the $\Lambda$CDM model usually predict a diverging central halo DM density
 $\rho_{DM}(r)\propto 1/r^\alpha,\alpha\st{>}{\sim}1$, where
$r$ is a distance from the center of a galaxy,
 while  observations indicate that some
low surface brightness galaxies have an almost flat core density.
(However, some dwarf spheroidal  satellites of the Milky Way have steeper DM density profiles.)
 The CDM theory also predicts many  satellite galaxies $(O(10^3))$ around the Milky Way, however, only $O(10^2)$ satellite
galaxies are observed at most
 ~\cite{Salucci:2002nc,navarro-1996-462,deblok-2002,crisis}.

Recently, it was also found that the mass enclosed within a radius
of $300~pc$ in these galaxies is approximately constant ($\sim
10^7 M_\odot$) regardless of their luminosity between $10^3$ and
$10^7 L_\odot$ ~\cite{Strigari:2008ib}. This result implies the
existence of a  minimum mass scale for
 the dwarf  galaxies~\cite{Mateo:1998wg,gilmore} independent of their  baryon matter fraction,
 in addition to the minimum length scale.
 This observational fact could be also problematic with the $\Lambda$CDM model,
which usually predicts many dark matter dominated structures smaller than
the dwarf galaxies down to
$10^{-6}M_\odot$~\cite{1475-7516-2005-08-003}.
( Strigari et al ~\cite{Strigari:2008ib} suggested that
this difficulty can be overcome if we consider roles of baryonic matter.)

 In this paper, we
suggest that a cold dark matter model based on  Bose-Einstein
condensate (BEC) ~\cite{sin1}
  or scalar field dark matter (SFDM)~\cite{myhalo}
  can  explain this minimum mass as well as the minimum length scale observed.
 Considering observational and theoretical uncertainties,
we are here concerned with the scales correct  within an order of magnitude.

Let us briefly review the history of the DM theory based on  BEC.
  In 1992, to explain the observed  galactic rotation curves, Sin \cite{sin1,sin2} suggested that galactic halos
are   astronomical objects in BEC  of
ultra-light (with mass $m\simeq  O(10^{-24})eV$) DM particles such as
pseudo Nambu-Goldstone boson.
 In this model the halos are like gigantic atoms, where
the ultra-light boson DM particles are condensated in a single macroscopic quantum state $\psi$.
 The quantum mechanical uncertainty principle
prevents halos from
 a self-gravitational collapse.

Lee and Koh ~\cite{kps,myhalo} suggested that  this  condensated halo DM can be described
as a coherent scalar field \cite{myreview} (dubbed as the boson halo model or,
more generally,  SFDM
model later ~\cite{0264-9381-20-20-201} ).
 On the other hand, in usual CDM models, the wave functions of DM particles
 do not overlap much and the particles move
 incoherently.
 Thus, the key difference between our model and the usual CDM models
 is the state of DM particles rather than DM itself.
 Later,  similar ideas were  suggested  by many authors
 in terms of  the fuzzy DM, the fluid DM, or the repulsive DM ~\cite {Schunck:1998nq,PhysRevLett.84.3037,PhysRevD.64.123528,PhysRevD.65.083514,
repulsive,Fuzzy,corePeebles,Nontopological,PhysRevD.62.103517,Alcubierre:2001ea,
Fuchs:2004xe,Matos:2001ps,0264-9381-18-17-101,PhysRevD.63.125016,Julien,moffat-2006,Akhoury:2008dv}.
(See ~\cite{DMmodels} for a  review.)
Many related models like monopole failing models~\cite{Matos:2000ki,Nucamendi:2000is},
unstable massless scalar field models~\cite{Schunck:1998nq}, and quintessence models~\cite{PhysRevD.64.123528}
are suggested too.
There are also extensive literature on, for example,  the collision of BEC-dark matter structures~\cite{Bernal:2006ci,mycollision},
 non-spherical collapse~\cite{Bernal:2006it} and virialization processes~\cite{Bernal:2006it},
  and the collapse of SFDM fluctuations in an expanding universe~\cite{Fuzzy,Guzman:2004wj,Fukuyama:2007sx,Matos:2009hf,guzman:024023}.
(For a review see \cite{myreview,Schunck:1998nq,Matos:2001ps} and references therein.)

In these models (BEC/SFDM model hereafter) the formation of DM
structures smaller than the Compton wavelength ($O(1)~ pc$) of DM
particles is suppressed by the quantum uncertainty principle.
(We will see that the actual length scale is somewhat larger.)
It
was shown that this property could  alleviate the problems
of the $\Lambda$CDM model
~\cite{corePeebles,PhysRevD.62.103517,0264-9381-17-13-101,PhysRevD.63.063506},
such as the cusp problem~\cite{Klein:2008mn} and missing satellite
problem. It is also suggested that this model can well explain the
observed rotation curves~\cite{Boehmer:2007um}, collisions of
galaxy clusters~\cite{mycollision} and the size evolution of the
massive galaxies~\cite{mysize}. Note that the BEC/SFDM particles
behave like CDM ~\cite{Matos:2003pe} at the scale larger
than galaxies during the cosmological structure formation due to a
small velocity dispersion of the Bose-Einstein  condensated state. Thus, this model safely satisfies the
criteria of the well established structure formation theory above
the galactic scales.

In this paper, we show that the minimum mass of galaxies can be explained
if DM is in BEC.
In Sec. 2 we show the relation
between the minimum mass and the minimum length scale of galaxies.
 In Sec. 3 we review the proposed origin of the minimum length scale in the BEC/SFDM model.
 In Sec. 4 we present a result of a simple numerical study supporting our theory.
 Section 5 contains discussion.

\section{Minimum mass of galaxies from the minimum length scale}

The characteristic length
scale ($O(1)\sim O(10^3)~pc$)  for the clustering of dark matter
is the key feature of the BEC/SFDM~\cite{sin1,myhalo} and its variants.
It is very interesting that this minimum length scale of DM halo seems to be already observed ~\cite{gilmore}.
There are no known stable galaxies with a half-light radius smaller than $120~pc$
while the maximum  size for star clusters deprived of DM is about $30~pc$.
Furthermore, observational data
indicate there is
a distinction   in phase-space density between a star-cluster and
 a dwarf galaxy~\cite{Walcher}.
 Since galaxies, especially dwarf galaxies, are highly DM dominated objects,
it is plausible  that the minimum mass of galaxies observed is
connected with DM rather than
 visible matter in the galaxies.
Thus, it is natural to think that  the minimum mass of galaxies is somehow related  to  this finite length scale.

First, let us briefly review the BEC/SFDM model, in which
a  galactic dark halo is described by a quantum wave function $\psi(r)$ (or scalar field) of
the non-linear Schr\"{o}dinger equation
with the Newtonian gravity~\cite{sin1};

\beq
i\hbar\partial_t \psi =E\psi =
-\frac{\hbar^2}{2m}\nabla^2\psi+
 V(r) \psi(r),
\label{sch} \eeq where $E$ is  energy of a DM particle and $r$
is the distance from the halo center. The gravitational potential of the halo
is given by
\beq
\label{Vr}
 V(r)= \int^{r}_0 dr'\frac{Gm}{r'^2}\int^{r'}_0
dr'' 4\pi r''^2 (\rho_{vis}(r'')+ \rho_{DM}(r'') ) +V_0, \label{V}
\eeq
where DM density $\rho_{DM}(r)=M_0 |\psi(r)|^2$ and
$\rho_{vis}$ is the visible matter (i. e., stars and gas) density.
$M_0$ is a mass parameter and $V_0$ is a constant to make
$V(\infty)=0$. These equations can be also derived in the SFDM theory
in the Newtonian limit. For simplicity, we assume there is no
self-interaction between the scalar particles except for the gravity.

Let us find the size of a stable configuration of the halo.
From Eq. (\ref{sch}) the energy $E$  is approximately given by
%
\beq
E(\xi) \simeq\frac{\hbar^2}{2m \xi^2}+\int^{\xi}_0 dr'\frac{Gm}{r'^2}\int^{r'}_0 dr'' 4\pi r''^2
(\rho_{vis}(r'')+ \rho_{DM}(r'') ),
\eeq
as a function of a halo length scale $\xi$.
A stable ground state configuration of the halo suitable for dwarf galaxies (a zero-node solution)
can be approximately found by
extremizing the energy by $\xi$~\cite{Silverman:May};
\beq
\label{dE}
dE(\xi)/d\xi\simeq -\frac{\hbar^2}{m \xi^{3}}+\frac{GM m}{\xi^2}=0.
\eeq
Here,
\beq
M\equiv \int^{\xi}_0 dr'' 4\pi r''^2
(\rho_{vis}(r'')+ \rho_{DM}(r''))
\eeq
 is the total mass within $\xi$ of the galaxy consisting $both$ of the
DM  and the  visible matter.
The condition in Eq. (\ref{dE}) satisfies at ~\cite{sin1,Silverman:May}
\beq
\label{xi}
\xi=\frac{\hbar^2}{GMm^2}=\frac{c^2\lambda_c^2}{4\pi^2 GM},
\eeq
where $c$ is the light velocity and $\lambda_c=2\pi \hbar/mc$ is the Compton wavelength
of the particles.
From this equation one can obtain the  mass within $\xi$,
\beq
\label{Mc}
M(\xi)\simeq\frac{\hbar^2}{G\xi m^2}.
\eeq

It is well known that, in the BEC/SFDM theory, the size of the stable configuration
$\xi$ could not be arbitrary small. Let us denote this minimum value of $\xi$ as $\xi_c$ which
is at least  $\lambda_c$. Then, there appears a minimum mass for the DM halo,
 $M_c\equiv M(\xi_c)={\hbar^2}/{G\xi_c m^2}$ corresponding to $\xi_c$.
If $\xi_c=\lambda_c\sim 1/m$
 and $m\simeq 10^{-24} eV$, the minimum mass for the DM halo becomes
 $M_c=\hbar c/2\pi G m\simeq 10^{13} M_\odot$, which is too large for dwarf galaxies.
However, note that the free parameter mass   $m\sim 10^{-24} eV$ is just one of  rough estimates.
In the  literature on the BEC/SFDM, $m$ was usually suggested to be in the range
$O(10^{-26})eV\sim O(10^{-22})eV$ to solve the problems of CDM models.

 Conversely,
from the observed values of  $M_c$ and $\xi_c$ one can obtain  approximate mass of DM particles
\beq
\label{m}
m\simeq \sqrt{\frac{\hbar^2}{G\xi_c M_c}}.
\eeq
 From the observed value $M_c\simeq 10^7 M_\odot$, one can obtain
$m\simeq 5.4\times 10^{-22}~eV$ for $\xi_c=300~pc$ from Eq. (\ref{m}).
For this $m$, $\lambda_c\simeq 0.075 pc$.
(In the next section we will discuss why $\xi_c$ is larger than $\lambda_c$.)
Very interestingly, this $m$ value is similar to the value $m= 10^{-22}eV\sim 10^{-23}eV$ required to solve the cusp problem
and to suppress the small-scale  power ~\cite{Fuzzy,Alcubierre:2001ea,Alcubierre:2002et}.
Thus, $m\sim 10^{-22} eV$ can solve all known small scale problems mentioned above, which
was discussed by Hu, Barkana and Gruzinov in Ref. ~\cite{Fuzzy}.

\section{Minimum length scale of galaxies revisited}

In fact, even in the original work ~\cite{sin1,myhalo}  it had been already
  noted that   the typical length scale ($O(kpc)$) of galaxies is somewhat larger than
  the Compton length $\lambda_c$.
  This is related to the fact that the halos are basically non-relativistic objects.
  $\xi_c$ could be comparable to $\lambda_c$ only when astronomical objects in consideration
   are extremely relativistic like black holes.
  It is clearly not the case for galactic halos.
 $\xi$ in our equation (\ref{xi}) was identified as the gravitational Bohr radius $\sim kpc$ in \cite{sin1}.
In \cite{myhalo} it was argued that de Broglie wavelength $\lambda_{dB}$, rather than the
   Compton length,
   is more adequate for
   the typical size of  DM structures.
   (Similarly, in condensed matter BEC systems the coherence length or the healing length,
usually comparable to  $\lambda_{dB}$, rather than $\lambda_c$
determines the spatial size of BEC density
fluctuations~\cite{BECreview}.)

 There are several  ways to
obtain $\xi_c$ slightly larger than $\lambda_c$.
The first and most plausible one  is the quantum
Jeans scale $r_J \sim O((G\rho m^2)^{-1/4})$ ~\cite{Sahni:1999qe,Fuzzy,Silverman:May}
  for SFDM, which
is the geometric mean between the
virial dynamical scale
and the Compton scale.
This scale can be derived from the cosmological evolution equation for the
 SFDM density perturbation $\delta$ with a wave vector $k$ ~\cite{Fuzzy,Sikivie}
\beq
\partial^2_t\delta+2H\partial_t \delta -\left( 4\pi G\rho -\frac{k^4}{4 m^2 a^4}\right) \delta=0,
\eeq
where $H$ is the Hubble parameter, $\rho$ is  the mean DM
density and $a$ is the scale factor. The second term in the parenthesis represents a
contribution from quantum pressure of the DM.
The density perturbations are
stable below $r_J$ and behave as ordinary CDM above this scale.
This fact makes the BEC/SFDM an ideal alternative to the CDM.
  $r_J$ determines
  the minimum length scale during the cosmic structure formation,
which could be  larger than $\lambda_c$.
(This quantum Jeans instability  is also known in the studies of boson stars
~\cite{Khlopov,PhysRevD.41.2998}.)
According to the standard structure formation theory, a physically meaningful value of $r_J$ can be
 fixed at the  matter-radiation equality having a scale factor $a_{eq}\simeq 1/3200$.
 At this time $\rho\simeq 3 \times 10^{10}
M_\odot  a_{eq}^{-3}/Mpc^3$~\cite{wmap3}, and the quantum Jeans scale $r_J$ above becomes
 $O(10^2) pc$ for $m\simeq 10^{-22}eV$. Hence, $\xi_c\simeq r_J$ is plausible.
Second, $\xi_c$ can be the coherence length  $\sqrt{M_P/\psi_0} \hbar/mc$
determined by the DM scalar field value $\psi_0$
at the halo center ~\cite{PhysRevD.65.083514}, which might be related to some symmetry breaking.
For a simple quadratic potential $\rho\sim m^2 \psi_0^2$,
the coherence length  again becomes of order  $r_J$ ~\cite{Silverman:May}.
 This implies $\psi_0\sim 10^{-8} M_P$, where $M_P$ is
the Planck mass.
Third, if there is a self-interaction term $\lambda \phi^4$ between DM particles,
a new length scale $\xi_c=O(\lambda^{1/2}\lambda_c M_P /m)$ emerges ~\cite{Schunck:1998nq}.
Finally, one may  also consider a thermal de Broglie wavelength
 for $\xi_c$~\cite{Silverman:May}.
Thus, $\xi_c$ somewhat larger than $\lambda_c$ is not new in these models and
theoretically possible.

We  assume one of these scales plays a role of $\xi_c$, though
the quantum Jeans scale  seems to be most
 plausible for our purpose.
Note that all these scales are determined not by the properties of
individual halos but by those of the DM particles such as mass, coupling and mean density of the DM particles.
Therefore,  $\xi_c\simeq O(10^2)~pc>~\lambda_c$ could be a universal quantity~\cite{gilmore}.
In short, the nature of the DM particles fixes $\xi_c$, which in turn decides $M_c$.

To make a DM dominated structure,
the total mass of the structure should be large enough to gravitationally attract DM particles having intrinsic
 momentum of $O(\hbar/\xi_c)$ due to the uncertainty principle.
 The critical mass $M_c$ is  just the minimum value to do this.
 From the virial condition, i. e., by equating kinetic energy ($(\hbar/\xi_c)^2/2m)$ with potential energy
 ($V= G M m/\xi_c$), one can check that $M_c$ in Eq. (\ref{Mc}) is of correct order.
It is important to note that, since $\xi_c$ is the property of DM itself, $M_c$ should be independent of the fraction of visible matter, hence the luminosity,
as long as the fraction is small.
This seems to be the basic physics behind what was observed.

The thermal history of this dark matter is also considered by several authors~\cite{sin3,Silverman:May,1475-7516-2004-12-012}.
Since the dark matter particle is ultra-light, its thermal velocity could be highly relativistic,
and some concerns about the BEC transition may arise.
There are basically two options. First, BEC/SFDM might simply have been generated by a non-thermal process
like   axions.
Second, even when the DM particles
are  in thermal equilibrium,
 contrary to common wisdom, a  BEC could form
for relativistic bosons~\cite{PhysRevLett.46.1497,PhysRevLett.66.683,grether-2007-99,1475-7516-2004-12-012}.
In~\cite{1475-7516-2004-12-012} it is shown that a cosmological BEC always exists
for the relativistic SFDM, if $m < 10^{-14} eV$~\cite{PhysRevLett.46.1497,PhysRevLett.66.683,grether-2007-99}.

After the BEC phase transition at the temperature $T_c$, almost all DM particles are in a ground state
and move coherently rather than randomly.
The BEC ground state are favored against  thermally excited states owing
 to the Bose-Einstein statistics of boson particles~\cite{Silverman:May}.
 According to ~\cite{1475-7516-2004-12-012} the present DM number density
 $n\sim 10^{15} eV^{3}$ is much larger than the maximum charge density allowed to excited states;
 $m T_0^2/3\sim 10^{-30} eV^{3}$,
even when the present temperature of the DM
  particles is as high as $T_0\sim 10^{-4}eV$ for $m\simeq 5.4\times 10^{-22}~eV$.
 (See Eq. (13) of the reference.)
 Hence, almost all BEC/SFDM particles are in the ground state now
 even in this case.

Once the relativistic BEC is formed,
 SFDM particles in the ground state  have only a small quantum velocity dispersion $\sim\hbar/m \xi$
 and could be bounded in a self-gravitational potential (if $M>M_c$)
 even though the temperature of the condensate could very
high compared to the mass of the particles.
The quantum velocity dispersion
$\Delta v\sim   h /m \xi_c\sim \lambda_c/\xi_c\sim 10^{-4} c\sim O(10)km/s$ is
smaller than the escape velocities of dwarf galaxies.
For stability analysis, we can think of the boson halos as boson stars~\cite{myhalo,Schunck:1998nq}.
Their stability
with respect to various kinds of perturbations or gravitational collapse is shown  in
many works~\cite{0264-9381-20-20-201}.
All these facts indicate the stability of the dark halos with the BEC/SFDM~\cite{bernal:063504}.
There are also universal profiles for boson stars, which are attractors of
 the collapse of quite arbitrary initial density fluctuations ~\cite{gcooling2}.

\section{A numerical simulation}

To be more concrete, we perform a numerical study using the shooting method~\cite{myhalo}
for a toy  DM halo.
Similar numerical work for cases without visible matter
has been extensively performed in the studies of boson stars~\cite{0264-9381-20-20-201}.
The dimensionless form of
  Eq. (\ref{sch}) and Eq. (\ref{Vr})
 can be  written as
 \beq
 \left\{ \begin{array}{l}
 \nabla _{\bar{r}} ^2 {\bar{V}} = (\sigma ^2+{\bar{\rho}}_{vis})  \\
  \nabla_{\bar{r}} ^2 \sigma  = 2{(\bar{V}-\bar{E})}\sigma    \\
\end{array} \right. .  \\
\label{all}
 \eeq
Here  we have introduced dimensionless quantities ~\cite{myhalo}
\beqa
\label{transform}
\sigma &\equiv& \sqrt{4\pi G}e^{-i\bar{E}\bar{t}}\psi/c^2, \no
\bar{r} &\equiv&   mc{r}/\hbar,   \no
 \bar{t} &\equiv& mc^2 t /\hbar
 \eeqa
  so that
all other barred quantities are dimensionless.
For example, in this unit, $\bar{\xi}\equiv  mc{\xi}/\hbar$,  $\bar{V}\equiv V/c^2$ and $\bar{E}\equiv E/mc^2$.

\begin{figure}[htbp]
\includegraphics[width=0.6\textwidth]{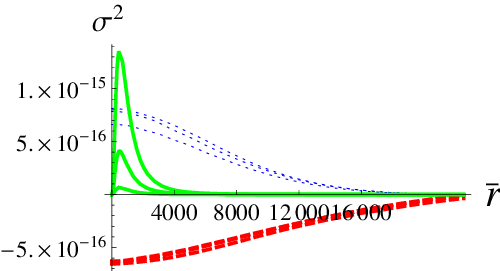}
\caption{ (Color online)
The dark matter density $\sigma^2$ (blue dotted lines),
the visible matter density  $\bar{\rho}_{vis}$ (green thick lines)   and
the gravitational potential $\bar{V}$ (red dashed lines, in units of $c^2$, scaled by $3\times 10^{-8}$)
for a model dwarf galaxy as functions of
distance  $\bar{r}$ from the halo center.
Despite of wide variations of a visible matter fraction, the 3 profiles have
a universal core size $\bar{r}_c=4000$ (corresponding to the physical size $r_c \simeq 300 pc$) and the total mass within
this size $\bar{M}_{tot} \simeq 0.00019 ~$ (corresponding to $4.75\times 10^7 M_\odot$) is similar for all 3 cases and the gravitational
 potential functions almost overlap. Here,
 $\bar{r}=1$ corresponds to a physical distance $1/m\sim 0.075~pc$ and $\sigma=1$ to the field value
    $|\psi|=1/\sqrt{4\pi G}$.  }
 \label{r}
\end{figure}

For visible matter of dwarf galaxies, we use
the dimensionless version of the empirical density in Ref. ~\cite{Brownstein:2005zz}
\beq
\bar{{\rho}}_{vis}=3 \beta  M_{vis} \left(\frac{\bar{r}}{\bar{r}_c+\bar{r}}\right)^{3\beta}
 \frac{\bar{r}_c}{4 \pi \bar{r}^3 (\bar{r} + \bar{r}_c)},
\eeq
where $\beta=2$ and $M_{vis}$ is a parameter proportional to the visible matter fraction.
It has been shown that this $\bar{\rho}_{vis}$ with the BEC/SFDM can successfully
reproduce the observed rotation curves of dwarf galaxies~\cite{Boehmer:2007um}.
We choose the visible core size $\bar{r}_c=664$
corresponding to the physical size  $r_c\simeq 40~pc$.
Considering the observational data, we choose  $\bar{\xi_c}=3985~(\xi_c=300~pc)$.
One can calculate a dimensionless total mass within $\bar{\xi}_c$,
\beq
\bar{M}_{tot}\equiv \int_0^{\bar{\xi}_c}
 4 \pi \bar{r}^2 (\sigma(\bar{r})^2 + \bar{\rho}_{vis}(\bar{r})) d\bar{r},
 \eeq
which corresponds to a physical mass $M_{tot}=\bar{M}_{tot} M_P^2/m$.

Fig. 1 shows the result of our numerical study with boundary conditions
$d\bar{V}/d\bar{r}(0)=0,\bar{V}(\infty)=0 $ and $d\sigma/d\bar{r}(0)=0$.
 For other parameters, we consider 3 cases with
the  parameters  $(M_{vis} = 5\times 10^{-6},\sigma(0) = 3.6\times 10^{-8}, \bar{V}(0)= -2.18\times 10^{-8},\bar{E}=-1.7\times 10^{-9}),
(M_{vis} = 3\times 10^{-5},\sigma(0) = 3.4\times 10^{-8}, \bar{V}(0)= -2.12\times 10^{-8},\bar{E}=-1.3\times 10^{-9})$, and $
(M_{vis} = 1\times 10^{-4},\sigma(0) =2\times 10^{-8}, \bar{V}(0)= -2.097\times 10^{-8}, \bar{E}=-1.28\times 10^{-9})$, respectively, from
the top to the bottom for $\sigma(0)$ (reversely for $\bar{\rho}_{vis}$).
( One can easily recover the dimensionful quantities by using the transformation in Eq. (\ref{transform}) and below.
For example, $\bar{E}=-1.7\times 10^{-9}$ corresponds to the real energy $E=\bar{E} mc^2=-9.18 \times 10^{-31}~eV$.)
Three profiles of the potential functions  almost overlap as our theory expected.
For a similar mass $\bar{M}_{tot}\simeq 0.00019$, the shapes of three $\bar{V}(r)$ are similar, regardless of $M_{vis}$, as long as
DM is dominant.
The converse is also true.

The result confirms our arguments below Eq. (\ref{Mc}).
Despite of wide variations of visible matter fraction, the profiles of DM and potential
well have
a universal core size ($\xi_c\simeq 300~pc$) and the total masses within $300 pc$,
 $M_{tot}=(0.000189,0.000190,0.000189)$, are similar for all 3 cases.
(In the physical units these are
$M_{tot}\simeq 0.00019 M_P^2/m\simeq 4.75\times 10^7 M_\odot \simeq  M_c$).
The masses of the visible matter are $(1.25\times 10^6,7.5\times 10^6,2.5\times 10^7)M_\odot$, respectively.
Thus, the densities of dark and visible matter
are similar to the  observed values.
In our simulation, what really matter were $M_c$ and $\xi_c$ not the composition of matter within $\xi_c$.
This confirms that there is
 a universal minimum mass of galaxies
 composed of the DM and ordinary matter
corresponding to the characteristic length scale in the BEC/SFDM theory.
This consistency of the theory and observations
 is non-trivial because the numerical solutions of Eq. (\ref{all}) are very sensitive to the boundary conditions.
On the contrary, the results are not sensitive to the form of $\bar{\rho}_{vis}$. We have checked that a similar
conclusion can be derived  for $\bar{\rho}_{vis}$ with $\beta=1$.
When the DM is subdominant, the universal minimum mass and size are not shown
clearly. This is also expected because the universality is a property of the DM not that of
visible matter.

In \cite{gcooling2}
it is shown that under isolation conditions the rotation curve would be Keplerian,
 unless an appropriate spatial scale is chosen.
 We have used the cut-off of
  the scalar field profile at the boundary $\bar{r}=\bar{r}_0=22581~$ (corresponding to about $1700~pc$)  so that the density of the field equals the present average
  DM density of the universe $\rho_0\sim 3\times 10^{10} M_\odot/Mpc^3$, which corresponds to
  $\sigma(r_0)\simeq 10^{-11}$ in the dimensionless unit.
  This is necessary, because
   the DM density of galaxies is usually non-zero even for large r.
With $\psi\sim 10^{-8} M_P$ and $m\sim 10^{-22}eV$, we can successfully reproduce the observed halo DM density
$m^2 \psi^2\sim 10^{-4} eV^4\sim 10^7 M_\odot/(100pc)^3$ for dwarf galaxies as well as $\xi_c\sim O(10^2 pc)$.
Since above the galactic scale, SFDM basically behaves as CDM~\cite{Fuzzy},  we think that the usual bottom-up
hierarchial merging process happened for clusters of galaxies.

\section{Discussion}

We have shown analytically and numerically that the BEC/SFDM theory can explain the three observed properties of
dwarf galaxies, i.e.,
the minimum length scale, the minimum mass scale, and their independence from
 the brightness. On the other hand, it is difficult
 to explain all these properties
 in a $single$ scenario in the CDM context,
 although there are proposed solutions to some of these problems~\cite{Li:2008fx,Maccio':2008qt}
 relying on   roles of baryon matter.
 Another merit of our approach is that the stable DM configurations are
 not much dependent on
 the complicated galaxy merging history, because the equilibrium condition forces
the halos to rearrange their DM distribution so that they  have a universal form regardless of their history.
This could explain the observed universality of the mass and density profile independent
of  possible merging history.

In conclusion, the BEC/SFDM  could be a compelling alternative to the standard CDM, because it  could not only solve some known problems of CDM model such as the cusp problem and the satellite
problem, but also has a possibility to explain  recent observational mysteries of galaxy evolution in a simple way,
thanks to its wave nature and the characteristic length scale. Further tests and
future observations may reveal whether CDM is in a BEC.

\section*{ }
This work was supported  by the topical research program
(2009-T-1) of Asia Pacific Center for Theoretical Physics.


\end{document}